\documentclass[aps,preprint,showpacs,nofootinbib,12pt]{revtex4}
\usepackage{graphicx}
\usepackage{epsfig}
\usepackage[dvips]{color}

\begin{document}

\title{Possible molecular states of $D^{(*)}D^{(*)}$ and $B^{(*)}B^{(*)}$ within the Bethe-Salpeter framework}

\author{ Hong-Wei Ke$^{1}$   \footnote{khw020056@tju.edu.cn}, Xiao-Hai Liu$^{1}$ \footnote{xiaohai.liu@tju.edu.cn}  and
         Xue-Qian Li $^{2}$  \footnote{lixq@nankai.edu.cn}}

\affiliation{
  $^{1}$ School of Science, Tianjin University, Tianjin 300072, China \\
  $^{2}$ School of Physics, Nankai University, Tianjin 300071, China
 }

\begin{abstract}
\noindent

Recently the LHCb collaboration reported a new exotic state $T^+_{cc}$ which
possesses $cc\bar u\bar d$ flavor structure. Since its mass is
very close to the threshold of $D^0D^{*+}$ (or $D^{*0}D^{+}$) and
its width is very narrow, it is inclined to conjecture that $T^+_{cc}$ is a molecular state
of $D^0D^{*+}$ (or $D^{*0}D^{+}$). In this paper we study the
possible molecular structures of $D^{(*)}D^{(*)}$ and $B^{(*)}B^{(*)}$
within the Bethe-Salpeter (B-S) framework. We employ one boson
exchange model to stand the interaction kernels in the B-S
equations. With reasonable input parameters we find the isospin
eigenstate $\frac{1}{\sqrt{2}}(D^0D^{*+}-D^{*0}D^{+})$ ($J^P=1^+$)
constitutes a solution, which supports the ansatz of $T^+_{cc}$ being a molecular state of
$D^0D^{*+}$ (or $D^{*0}D^{+}$). With the same parameters we also
find that the isospin-1 state
$\frac{1}{\sqrt{2}}(D^{*0}D^{*+}+D^{*0}D^{*+})$ ($J^P=0^+$) can exist.
Moreover, we also study the systems of $B^{(*)}B^{(*)}$ and their
counterparts exist as possible molecular states. Consistency of theoretical computations
based on such states with the data of the
future experiments may consolidate the molecular structure of the exotic state
$T^+_{cc}$.

\end{abstract}


\maketitle

\section{introduction}
Recently the LHCb Collaboration reported a new exotic state $T_{cc}^+$
in the $D^0D^0\pi^+$ invariant mass spectrum. Apparently
$T_{cc}^+$ possesses a $cc\bar u \bar d$ flavor component. The
difference between its mass and the mass threshold of $D^0D^{*+}$ is
$-273\pm 61\pm5^{+11}_{-14}$ keV and its width is
$410\pm165\pm43^{+18}_{-38}$ keV\cite{LHCb:2021auc,LHCb:2021vvq}. The new state
attracts great interest because it is the first
double-charm tetraquark which was measured. Since 2003 many
exotic
states\cite{Choi:2003ue,Abe:2007jn,Choi:2005,Choi:2007wga,LHCb:2021uow,Collaboration:2011gj,LHCb:2019kea}
such as $X(3872)$, $X(3940)$,
$Y(3940)$, $Z(4430)^{\pm}$, $Z_{cs}(4000)$, $Z_{cs}(4220)$, $Z_b$,
$Z_b'$, $P_c(4312)$, $P_c(4440)$, $P_c(4457)$ have been measured. $T_{cc}^+$ have been observed in experiments and
the achievements broaden
our field of view about the flavor structure of exotic states.

It is difficult to attribute so many exotic states into a basket determined by the
traditional hadronic picture where a meson contains a quark and anti-quark pair and a baryon is composed of three valence
quarks. Instead, it is  suggested that they should be multi-quark states which are
predicted by the $SU(3)$ quark model\cite{GellMann:1964nj}.
During these years it turns out to be a hot topic to discuss the structures of
exotic states. Those new states are often proposed to be molecular
states, compact tetraquarks, a mixing of both structures or
dynamical effect\cite{Chen:2016spr,Guo:2019twa}.

The mass of $T_{cc}^+$ is very close to the mass threshold of
$D^0D^{*+}$ or ($D^+D^{*0}$) so naturally many authors suggested
that $T_{cc}^+$ could be a loose $D^0D^{*+}$ ($D^+D^{*0}$) bound
state\cite{Meng:2021jnw,Yan:2021wdl,Ren:2021dsi,Du:2021zzh,Xin:2021wcr,Feijoo:2021ppq}. Some
authors also consider it as a tetraquark\cite{Weng:2021hje,Agaev:2021vur}.
Generally a compact tetraquark has a wide decay width whereas a
molecular state has a relatively narrow one. Viewing the width of
$T_{cc}^+$ we also tend to accept $T_{cc}^+$ as a $D^0D^{*+}$
($D^+D^{*0}$) bound state. In this paper we study the possible
bound state of $D^0D^+$, $D^0D^{*+}$ and $D^{*0}D^{*+}$ systems
within the Bethe-Salpeter (B-S) framework where the relativistic
corrections are automatically included.

In this work we will study the systems composed of two charmed or
bottomed mesons and the scenario has not been explored in the B-S
framework yet. In our early
papers\cite{Ke:2012gm,Ke:2020eba,Ke:2019bkf} we deduced the B-S
equations for the systems containing  one vector and one
pseudoscalar, two pseudoscalars and two vectors respectively.
Following the approach in \cite{Ke:2012gm,Ke:2020eba,Ke:2019bkf}
we investigate the systems with two charmed or bottomed
constituents such as $T_{cc}^+$. Since the interaction kernels are
not the same as given in \cite{Ke:2012gm,Ke:2020eba,Ke:2019bkf}
and the objects under investigation are new it needs to re-study
the whole scenario.

If the interaction between two constituents is attractive and
large enough a bound state could be formed. In this work we employ
the one-boson-exchange model to calculate the interaction kernels
where the effective vertices are taken from the heavy meson chiral
perturbation
theory\cite{Colangelo:2005gb,Colangelo:2012xi,Ding:2008gr,Casalbuoni:1996pg,Casalbuoni:1992gi,Casalbuoni:1992dx}
. The exchanged particles are some light mesons such as $\pi$,
$\rho$ and $\omega$. We ignore the contribution from $\eta$
exchange because its mass is larger than $\pi$ and there exists an
additional suppression factor $\frac{1}{\sqrt{3}}$ at the
effective vertex (See Appendix A). In Ref. \cite{Ding:2008gr} the
authors indicated that $\sigma$ exchange makes a secondary
contribution, thus we also do not include it. With the effective
interactions we derive the kernel and establish the corresponding
B-S equation. The B-S equation is solved in momentum space so the
kernel we obtain by calculating the corresponding Feynman diagrams
can be used directly rather than converting it into a potential
form in coordinate space.

With all the input parameters, these B-S equations are solved
numerically. In some cases there no solution which satisfies the
equation exists as long as
the parameters are  set
within a reasonable range, it implies
the proposed bound state should not emerge in the nature. On the
contrary,  a solution of the B-S equation with reasonable
parameters implies that the corresponding bound state is formed.
In that case, the obtained B-S wave function can be used to calculate the decay rate of the bound state.

After this introduction we deduce the B-S equations and the corresponding kernels for the two
meson systems with different quantum numbers. Then in section III we
present our numerical results of the binding energies along with
explicitly displaying all input parameters. Section IV is devoted
to a brief summary.

\section{The Bethe-Salpeter formalism}

Initially, people employed the B-S equation to explore the bound
states of two fermions. Later this approach was extended to study
the bound states made of one fermion and one
boson\cite{Guo:1998ef,Weng:2010rb,Li:2019ekr}.  In
Refs.\cite{Guo:2007mm,Feng:2011zzb,Ke:2018jql,Feng:2012zzf,Ding:2021igr} the
B-S equation was used to study the spectra of the meson-meson
molecular states and then deal with their decays. The method was
extended to explore some other systems in our early
papers\cite{Ke:2012gm,Ke:2020eba,Ke:2019bkf,Ke:2021iyh}.

In Ref.\cite{Guo:2007mm,Feng:2011zzb} the B-S equation for a bound
state made of two pseudoscalars was deduced. Later we deduced the
B-S equations for a system composed of one pseudoscalar and one
vector or two vectors which are one particle and one
antiparticle\cite{Ke:2012gm,Ke:2020eba,Ke:2019bkf}.

In this work we are only concerned with the ground states where the orbital
angular momentum between two constituent mesons is zero (i.e. $l=0$).
For a system whose constituents are two pseudoscalars or one
pseudoscalar and one vector, its $J^{P}$ is $0^+$ or $1^+$. For
the molecular states which consist of two vector mesons their
$J^{P}$ may be $0^+$, $1^+$ and $2^+$.

Obviously, these systems composed of two charmed (or bottomed)
hadrons (off-shell ) should belong to the same representations of
isospin. In this case, the total wave function for the combined
systems of $D^0$ and $D^{+}$ ($D^{*0}$ and $D^{*+}$) must be
symmetric under group $O(3)\times SU_I(2)\times SU_S(2)$, where
$SU_I(2)$ and $SU_S(2)$ are isospin and spin groups respectively.
For the $D^0D^{+}$ system its total spin is 0 so its isospin
should be 1. Instead, for the $D^{*0}D^{*+}$ system  its isospin
is 0 as $J^{P}=1^+$, whereas it is 1 as $J^{P}=0^+$ or
$J^{P}=2^+$. For the $D^{*+}D^0$ or $D^{*0}D^+$ systems two
isospin states are possible:
$\frac{1}{\sqrt{2}}(D^{*+}D^0+D^{*0}D^+)$  ($I=0$) and
$\frac{1}{\sqrt{2}}(D^{*+}D^0-D^{*0}D^+)$ ($I=1$).

\subsection{The B-S equation of $0^+$ which is composed of two pseudoscalars}

The B-S wave function for the bound state $|S\rangle$ of two
pseudoscalar  mesons can be defined  as following:
\begin{eqnarray}\label{definition-BS1} \langle 0 | {\rm
T}\,\phi_1(x_1)\phi_2(x_2) | S \rangle = {\chi}_{{}_S}^{}(x_1,x_2)\,,
\end{eqnarray}
where $\phi_1(x_1)$ and $\phi_2(x_2)$ are the field operators of two
mesons, respectively, the relative coordinate $x$ and the center of mass coordinate
$X$ are
\begin{eqnarray} X=\eta_1 x_1 + \eta_2 x_2\,,\quad x = x_1 -
x_2 \,, \end{eqnarray} where $\eta_i = m_i/(m_1+m_2)$ and $m_i\,
(i=1,2)$ is the mass of the $i$-th constituent meson.

After some manipulations we obtain the B-S equation in the momentum space
\begin{eqnarray}\label{bs-equation-momentum1}
{\chi}_{{}_S}^{}(p) = \Delta_1\int {d^4p'\over
(2\pi)^4} { K_{S}}(p,p'){\chi}_{{}_S}^{}(p')\Delta_{2}\,,
 \end{eqnarray}
where $\Delta_i$ is the propagator of the $i$-th meson
and $ \Delta_1=\frac{i}{p_1^2-m_1^2}$, $ \Delta_2=\frac{i}{p_2^2-m_2^2}$.

The relative momenta and the total
momentum of the bound state in the equations are defined as
\begin{eqnarray} p = \eta_2p_1 -
\eta_1p_2\,,\quad p' = \eta_2p'_1 - \eta_1p'_2\,,\quad P = p_1 +
p_2 = p'_1 + p'_2 \,, \label{momentum-transform1}
\end{eqnarray}
where $P$ denotes the total momentum of the bound
state.

\begin{center}
\begin{figure}[htb]
\begin{tabular}{cc}
\scalebox{0.5}{\includegraphics{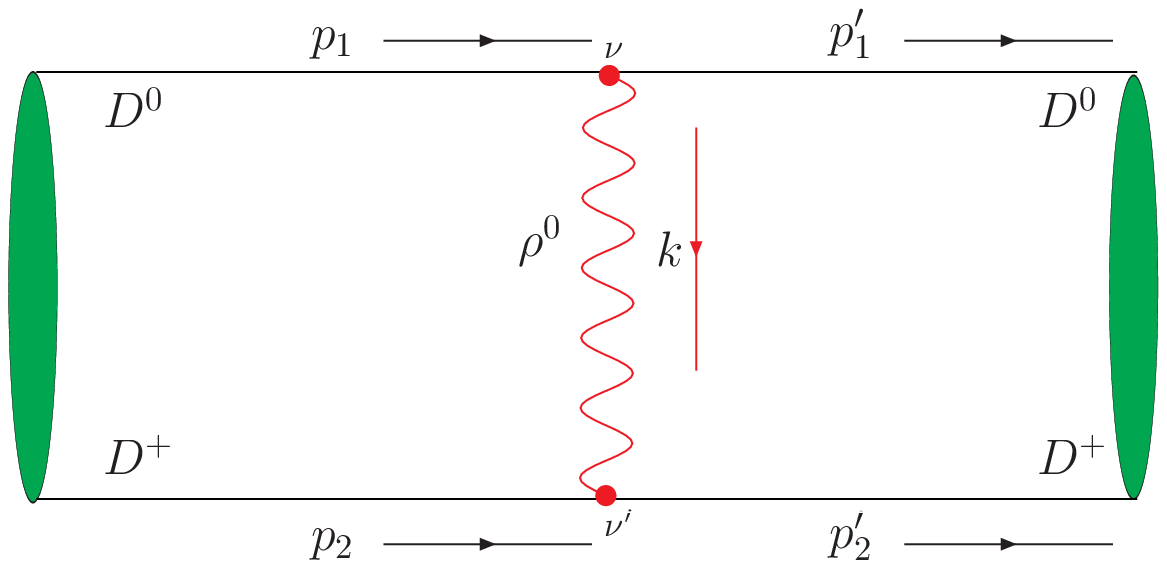}}\,\,\,\,\,\,\,\,\,\,\,\,\,\,\,\,\,\,\,\,\,\,\,\,\,\,\,\,\,\,\,\,\,\,\,\,\,\,\,\,\,\,\,\,\,\,\,\,\,\,\scalebox{0.5}{\includegraphics{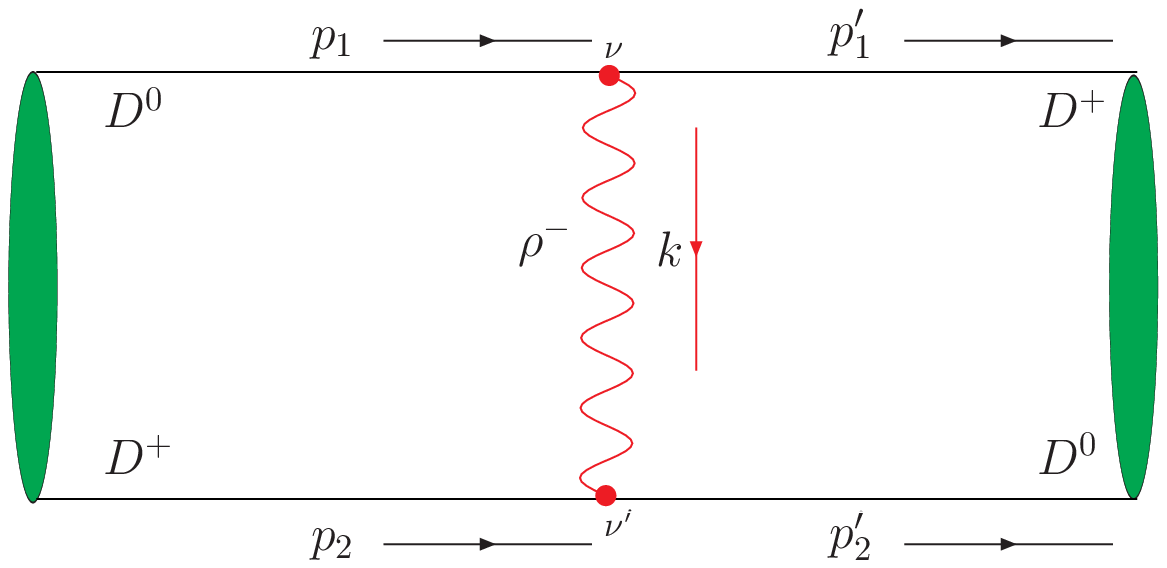}}\\
(a)\,\,\,\,\,\,\,\,\,\,\,\,\,\,\,\,\,\,\,\,\,\,\,\,\,\,\,\,\,\,\,\,\,\,\,\,
\,\,\,\,\,\,\,\,\,\,\,\,\,\,\,\,\,\,\,\,\,\,\,\,\,\,\,\,\,\,\,\,\,\,\,\,
\,\,\,\,\,\,\,\,\,\,\,\,\,\,\,\,\,\,\,\,\,\,\,\,\,\,\,\,\,\,\,\,\,\,\,\,
\,\,\,\,\,\,\,\,\,\,\,\,\,\,\, (b)
\end{tabular}
\caption{ A bound state composed of two pseudoscalars. In the
Feynman diagram (a)  $\omega$  exchange should also be included.}\label{DM21}
\end{figure}
\end{center}

 $D^0$ and $D^+$  can constitute two possible bound states : $\frac{1}{\sqrt{2}}(D^0D^++D^+D^0)$
with $I=1$ and $\frac{1}{\sqrt{2}}(D^0D^+-D^+D^0)$ with $I=0$.
Since only $l=0$ is considered and the total spin wavefunction is
symmetric the system of  an isospin-scalar is forbidden and the
isospin-1 state is reduced to $D^0D^+$. The exchanged mesons
between the two pseudoscalars are vector mesons, obviously we only
need to keep the lightest vector mesons $\rho$ and $\omega$
\cite{Guo:2007mm,Feng:2011zzb} ,  the Feynman diagrams
corresponding these effective interactions are depicted in Fig.
\ref{DM21}.

With the Feynman diagrams depicted in Fig. \ref{DM21} and the effective
interactions shown in appendix A we obtain the interaction kernel
\begin{eqnarray}
&&K_{S}(p,p')=K_{S0}(p,p',m_\rho)+{
K_{S0}}(p.p',m_\omega),\nonumber\\&&{K_{S0}}(p,p',m_{ V}) = i
 C_{S0} \, g_{DDV}^2
{(p_1+p_1')\cdot(p_2+p_2')-(p_1+p_1')\cdot q(p_2+p_2')\cdot q/m_{
V}^2 \over q^2-m_{ V}^2}F( q)^2,
\end{eqnarray}
where $q=p_1-p_1'$. For exchanging $\rho$ the expression
$K_{S0}(p, p',m_\rho)$ includes the contributions from figures
Fig. \ref{DM21} (a) and (b) but for exchanging $\omega$ it only includes the
contribution from figure Fig. \ref{DM21} (a). $C_{S0}=\frac{1}{2}$ for $\rho$
and $\omega$. Since the constituent meson is not a point particle,
a form factor at each interaction vertex among hadrons must be
introduced to reflect the finite-size effects of these hadrons.
The form factor is assumed to be in the following form:
\begin{eqnarray} \label{form-factor} F({k}) = {\Lambda^2 -
M_{ V}^2 \over \Lambda^2 -{k}^2}\,,
\end{eqnarray} where $\Lambda$ is a cutoff parameter.

Solving the Eq.(\ref{bs-equation-momentum1}) is
rather difficult. In general one needs to use the so-called
instantaneous approximation:$p_0'=p_0=0$ for ${K_{0}}(p,p')$  by which
the B-S equation can be reduced to
\begin{eqnarray} \label{3-dim-BS1}
{E^2-(E_1+E_2)^2\over (E_1+E_2)/E_1E_2}
\mathcal{\psi}_{{}_S}^{}({\bf p}) ={i\over
2}\int{d^3\mathbf{p}'\over(2\pi)^3}\, {\overline{} K_{S}}({\bf
p},{\bf p}')\mathcal{\psi}_{{}_S}^{}({\bf p}')\,,
\end{eqnarray}
where $E_i \equiv \sqrt{{\bf p}^2 + m_i^2}$, $E=P^0$, and the
equal-time wave function is defined as $ \psi_{_S}({\bf p})= \int
dp^0 \, \chi_{_S}(p) \,. $ For  exchange of a light vector between
the mesons, the kernel is
\begin{eqnarray}K_{S}({\bf p},{\bf
p}')=K_{S0}({\bf p},{\bf p}',m_\rho)+K_{S0}({\bf p},{\bf
p}',m_\omega),\end{eqnarray}
where the expressions of $K_{S0}(\mathbf{p},\mathbf{p}',m_V)$ can be found  in Appendix B.

\subsection{The B-S equation of $1^+$ which is composed of a pseudoscalar and a vector}
The B-S wave function for the bound state $|V\rangle$ composed of
one pseudoscalar and one vector mesons is defined as following:
\begin{eqnarray}\label{definition-BS2} \langle 0 | {\rm
T}\,\phi_1(x_1)\phi^\mu_2(x_2) | V \rangle ={\chi}_{{}_V}^{}(x_1,x_2)\epsilon^\mu,
\end{eqnarray}
where $\epsilon$ is the polarization vector of the bound state, ${\chi}_{{}_V}$ is the B-S wave function, $\phi_1(x_1)$ and $\phi^\mu_2(x_2)$ are respectively the
field operators of the two  mesons. The equation for the B-S wave
function is
\begin{eqnarray}\label{bs-equation-momentum2}
{\chi_{_V}}(p)\epsilon^\mu = \Delta_1 \int {d^4p'\over (2\pi)^4} {
K_{V\alpha\beta}}({ p},{
p}'){\chi_{_V}}(p')\epsilon^\beta\Delta_{2\mu\alpha}\,.
\end{eqnarray}
Here $ \Delta_1=\frac{i}{p_1^2-m_1^2}$ and $
\Delta_{2\mu\alpha}=\frac{i}{p_2^2-m_2^2}(\frac{p_{2\mu}p_{2\alpha}}{m_2^2}-g_{\mu\alpha})$
are the propagators of pseudoscalar and vector mesons. We
multiply an $\epsilon^*_\mu$ on both sides, sum over the
polarizations and  then deduce a new equation
\begin{eqnarray}\label{bs-equation-momentum2pp}
{\chi}_{{}_{V}}^{}(p) = \frac{-1}{3(p_1^2-m_1^2)(p_2^2-m_2^2)}\int
{d^4p'\over (2\pi)^4} {
K_{V\alpha\beta}}(p,p'){\chi}_{{}_{V}}^{}(p')(\frac{p_{2}^\mu
p_{2}^\alpha}{m_2^2}-g^{\mu\alpha})(\frac{P_\mu
P^\beta}{M^2}-g_\mu^\beta)\,.
 \end{eqnarray}

\begin{center}
\begin{figure}[htb]
\begin{tabular}{cc}
\scalebox{0.5}{\includegraphics{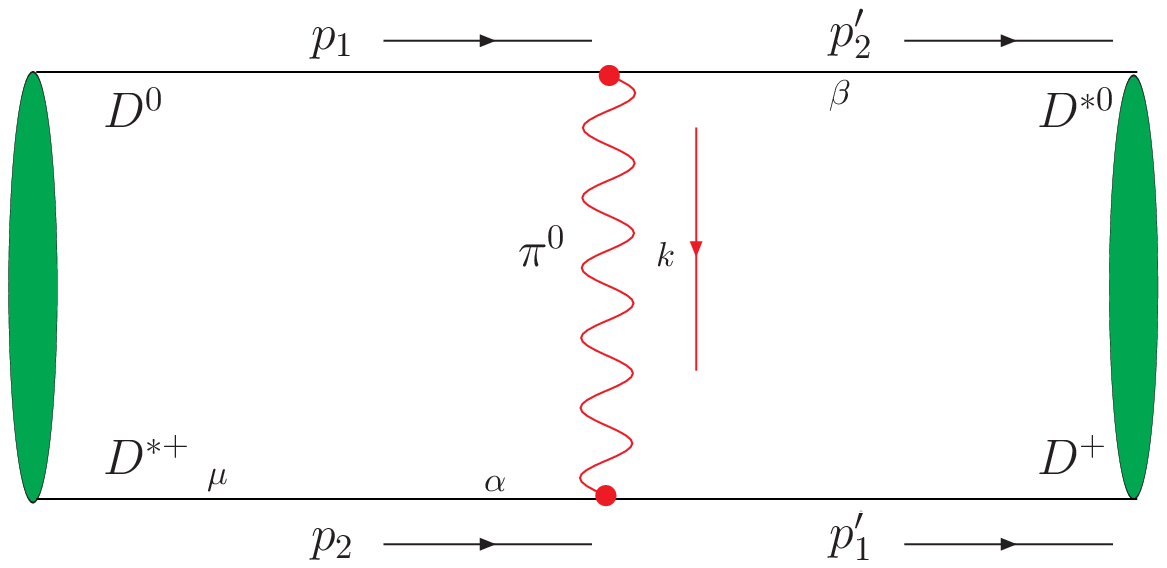}}\,\,\,\,\,\,\,\,\,\,\,\,\,\,\,\,\,\,\,\,\,\,\,\,\,\,\,\,\,\,\,\,\,\,\,\,
\,\,\,\,\,\,\,\scalebox{0.5}{\includegraphics{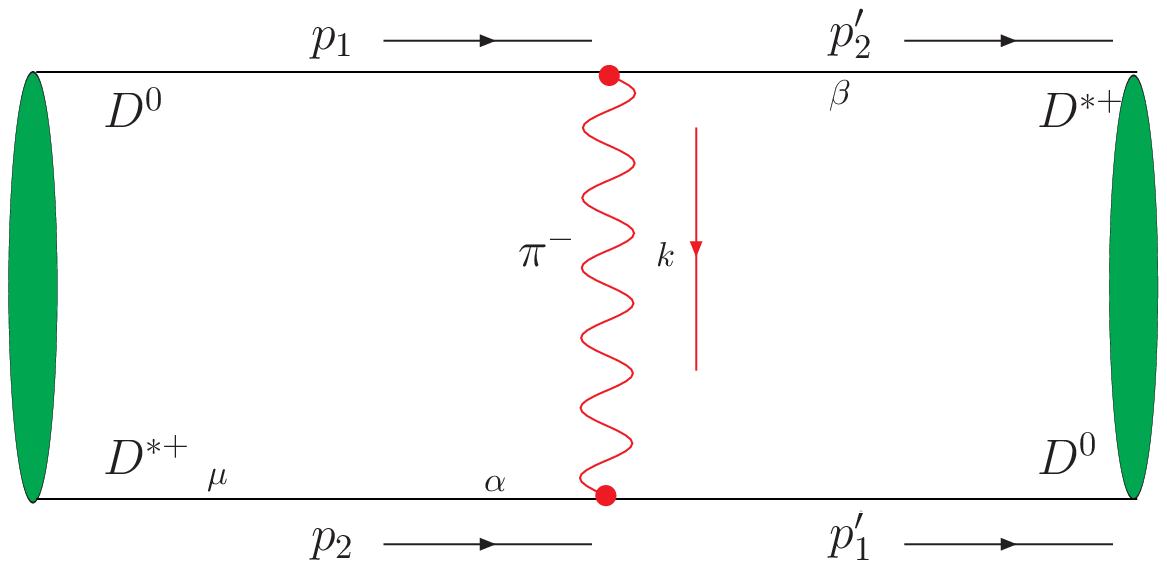}}\\
(a)\,\,\,\,\,\,\,\,\,\,\,\,\,\,\,\,\,\,\,\,\,\,\,\,\,\,\,\,\,\,\,\,\,\,\,\,
\,\,\,\,\,\,\,\,\,\,\,\,\,\,\,\,\,\,\,\,\,\,\,\,\,\,\,\,\,\,\,\,\,\,\,\,
\,\,\,\,\,\,\,\,\,\,\,\,\,\,\,\,\,\,\,\,\,\,\,\,\,\,\,\,\,\,\,\,\,\,\,\,
\,\,\,\,\,\,\,\,\,\,\,\,\,\,\, (b)
\end{tabular}
\caption{ a bound state composed of a pseudoscalar and a vector by
exchanging $\pi$.}\label{DM22}
\end{figure}
\end{center}

\begin{center}
\begin{figure}[htb]
\begin{tabular}{cc}
\scalebox{0.5}{\includegraphics{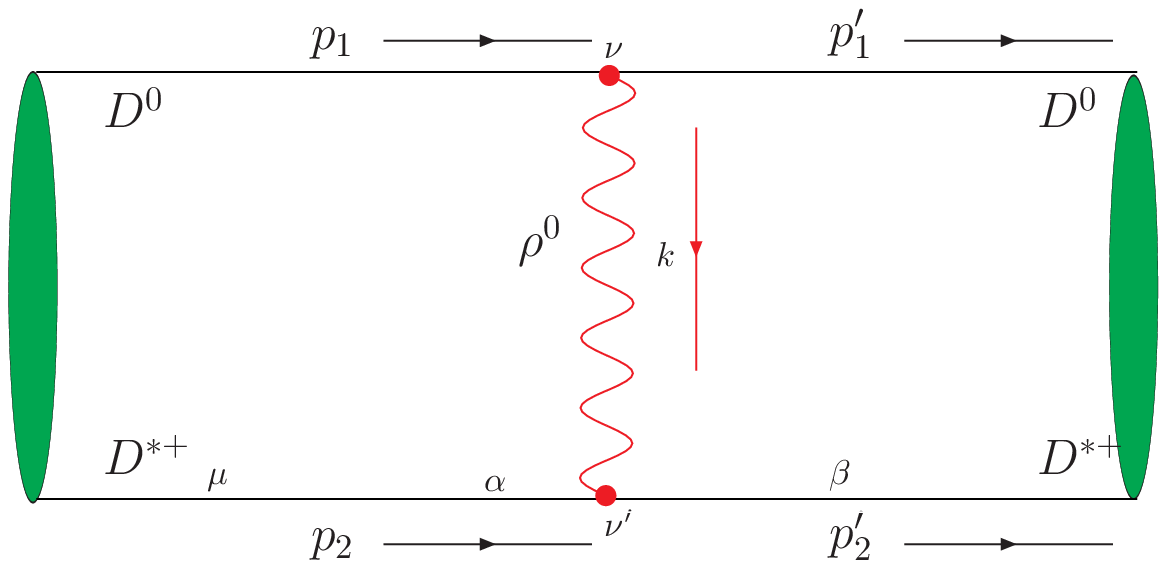}}\,\,\,\,\,\,\,\,\,\,\,\,\,\,\,\,\,\,\,\,\,\,\,\,\,\,\,\,\,\,\,\,\,\,\,\,
\,\,\,\,\,\,\,\scalebox{0.5}{\includegraphics{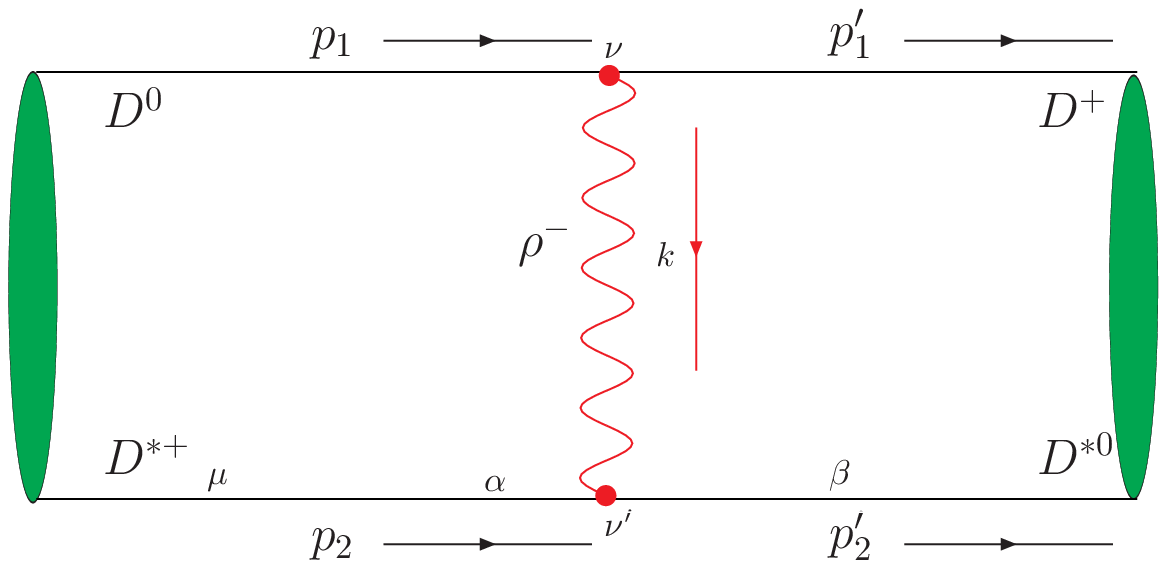}}\\
(a)\,\,\,\,\,\,\,\,\,\,\,\,\,\,\,\,\,\,\,\,\,\,\,\,\,\,\,\,\,\,\,\,\,\,\,\,
\,\,\,\,\,\,\,\,\,\,\,\,\,\,\,\,\,\,\,\,\,\,\,\,\,\,\,\,\,\,\,\,\,\,\,\,
\,\,\,\,\,\,\,\,\,\,\,\,\,\,\,\,\,\,\,\,\,\,\,\,\,\,\,\,\,\,\,\,\,\,\,\,
\,\,\,\,\,\,\,\,\,\,\,\,\,\,\,
(b)\\\scalebox{0.5}{\includegraphics{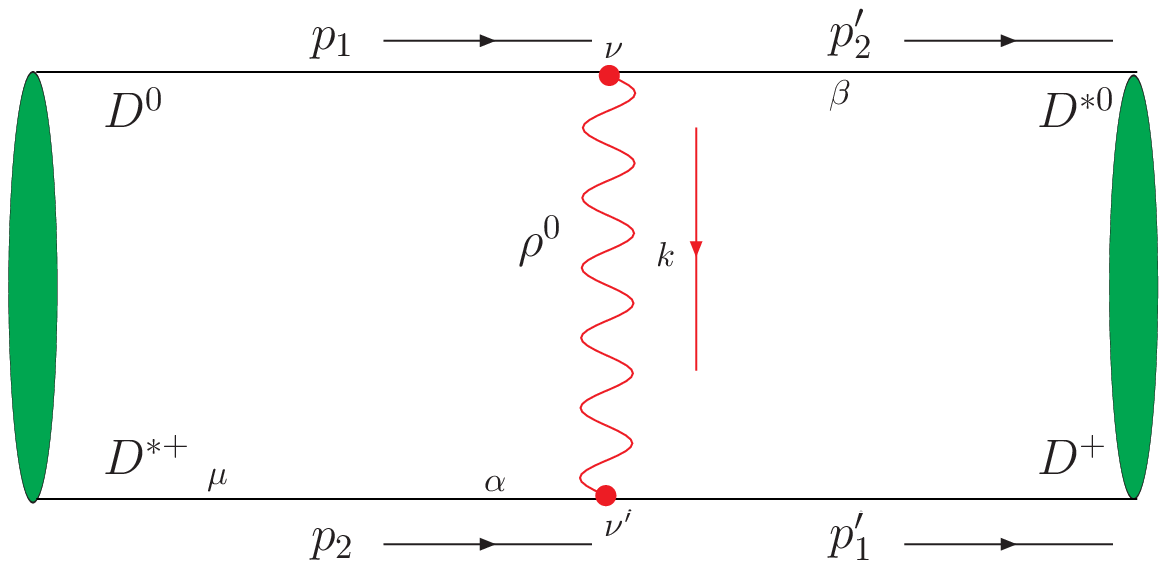}}\,\,\,\,\,\,\,\,\,\,\,\,\,\,\,\,\,\,\,\,\,\,\,\,\,\,\,\,\,\,\,\,\,\,\,\,
\,\,\,\,\,\,\,\scalebox{0.5}{\includegraphics{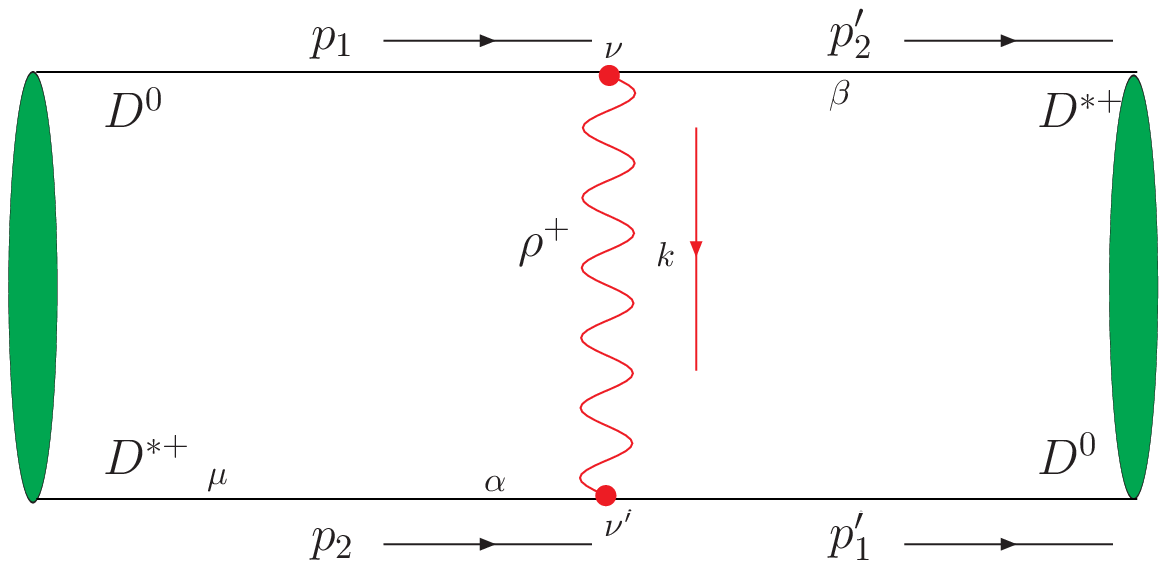}}\\
(c)\,\,\,\,\,\,\,\,\,\,\,\,\,\,\,\,\,\,\,\,\,\,\,\,\,\,\,\,\,\,\,\,\,\,\,\,
\,\,\,\,\,\,\,\,\,\,\,\,\,\,\,\,\,\,\,\,\,\,\,\,\,\,\,\,\,\,\,\,\,\,\,\,
\,\,\,\,\,\,\,\,\,\,\,\,\,\,\,\,\,\,\,\,\,\,\,\,\,\,\,\,\,\,\,\,\,\,\,\,
\,\,\,\,\,\,\,\,\,\,\,\,\,\,\, (d)
\end{tabular}
\caption{ A bound state composed of a pseudoscalar and a vector by
exchanging $\rho$. In the two Feynman diagrams (a) and (c)
exchange of $\omega$ also is also taken into account.}\label{DM23}
\end{figure}
\end{center}

With the Feynman diagrams depicted in Fig. \ref{DM22} and Fig.
\ref{DM23} we eventually obtain
\begin{eqnarray}
K_{V\alpha\beta}(p,p')&&=K_{V1\alpha\beta}(p,p',m_\rho)+{
K_{V2\alpha\beta}}(p,p',m_\rho)+{
K_{V1\alpha\beta}}(p,p',m_\omega)\nonumber\\&&+{
K_{V2\alpha\beta}}(p,p',m_\omega)+{
K_{V3\alpha\beta}}(p,p',m_\pi),\nonumber\\{K_{V1\alpha\beta}}(p,p',m_{
V})
&&=C_{V1}g_{_{DDV}}\{g_{_{D^*D^*V}}[g_{\alpha\beta}(p_2+p_2')\cdot(p_1+p_1')+g_{\alpha\beta}\frac{q\cdot(p_1+p_1')q\cdot(
p_2+p_2')}{m_{V}^2}] \nonumber\\&&+2g'_{_{D^*D^*V}}[q^\alpha
(p_1+p_1')^\beta-q^\beta
(p_1+p_1')^\alpha]\}\frac{i}{q^2-m_{V}^2}F(q)^2\nonumber\\{K_{V2\alpha\beta}}(p,p',m_{
V})&& = C_{V2}\varepsilon^{\mu\nu\beta\tau}(q'_\nu
g_{\lambda\mu}-q'_\mu
g_{\lambda\nu})(p_2'-p_1)_\tau\varepsilon^{\mu'\nu'\alpha\tau'}(q'_\mu
g_{\lambda'\nu}-q'_\nu
g_{\lambda'\mu})(p_2-p_1')_{\tau'}\nonumber\\&&\frac{i}{q'^2-m_{V}^2}(-g^{\lambda\lambda'}+q'^\lambda
q'^{\lambda'}/m^2_V)F(q')^2,\nonumber\\K_{V3\alpha\beta}(p,p',m_P)&&=C_{V3}g_{_{DD^*P}}^2\frac{i}{q'^2-m_{P}^2}q'_\alpha
q'_\beta F(q')^2,
\end{eqnarray}
where $q'=p_1-p_2'$. The contributions from Fig. \ref{DM22} are
included in $K_{V3\alpha\beta}(p,p',m_\pi)$  and those from Fig. \ref{DM23} (a)
and (b) are included in $K_{V1\alpha\beta}(p,p',m_V)$.
 When the bound state is an
isospin-scalar $C_{V1}=-\frac{3}{2}$ and  $C_{V2}=\frac{3}{2}$ for
$\rho$, $C_{V1}=\frac{1}{2}$ and  $C_{V2}=-\frac{1}{2}$ for
$\omega$ and $C_{V3}=\frac{3}{2}$ for $\pi$. When the bound state
is an isospin-vector $C_{V1}=\frac{1}{2}$ and $C_{V2}=\frac{1}{2}$
for $\rho$, $C_{V1}=\frac{1}{2}$ and $C_{V2}=\frac{1}{2}$ for
$\omega$ and $C_{V3}=\frac{1}{2}$ for $\pi$.

Defining $ K_{V}(p,p')={ K_{V\alpha\beta}}(q)(\frac{p_{2}^\mu
p_{2}^\alpha}{m_2^2}-g^{\mu\alpha})(\frac{P_\mu
P^\beta}{M^2}-g_\mu^\beta)$ and setting $p_0=q_0=0$ we derive the BS
equation which is similar to Eqs. (\ref{3-dim-BS1}) but possesses
a different kernel,
\begin{eqnarray} \label{3-dim-BS2}
{E^2-(E_1+E_2)^2\over (E_1+E_2)/E_1E_2}
\mathcal{\psi}_{{}_V}^{}({\bf p}) ={i\over
2}\int{d^3\mathbf{p}'\over(2\pi)^3}\, {\overline{} K_{V}}({\bf
p},{\bf p}')\mathcal{\psi}_{{}_V}^{}({\bf p}')\,,
\end{eqnarray}
where
\begin{eqnarray}
K_{V}(\mathbf{p},\mathbf{p}')&&=K_{V1}(\mathbf{p},\mathbf{p}',m_\rho)+{
K_{V2}}(\mathbf{p},\mathbf{p}',m_\rho)+{
K_{V1}}(\mathbf{p},\mathbf{p}',m_\omega)\nonumber\\&&+{
K_{V2}}(\mathbf{p},\mathbf{p}',m_\omega)+{
K_{V3}}(\mathbf{p},\mathbf{p}',m_\pi),
\end{eqnarray}
where the expressions of $K_{V1}(\mathbf{p},\mathbf{p}',m_V)$,
$K_{V2}(\mathbf{p},\mathbf{p}',m_V)$ and ${
K_{V3}}(\mathbf{p},\mathbf{p}',m_P)$ can be found  in Appendix B.

\begin{center}
\begin{figure}[htb]
\begin{tabular}{cc}
\scalebox{0.5}{\includegraphics{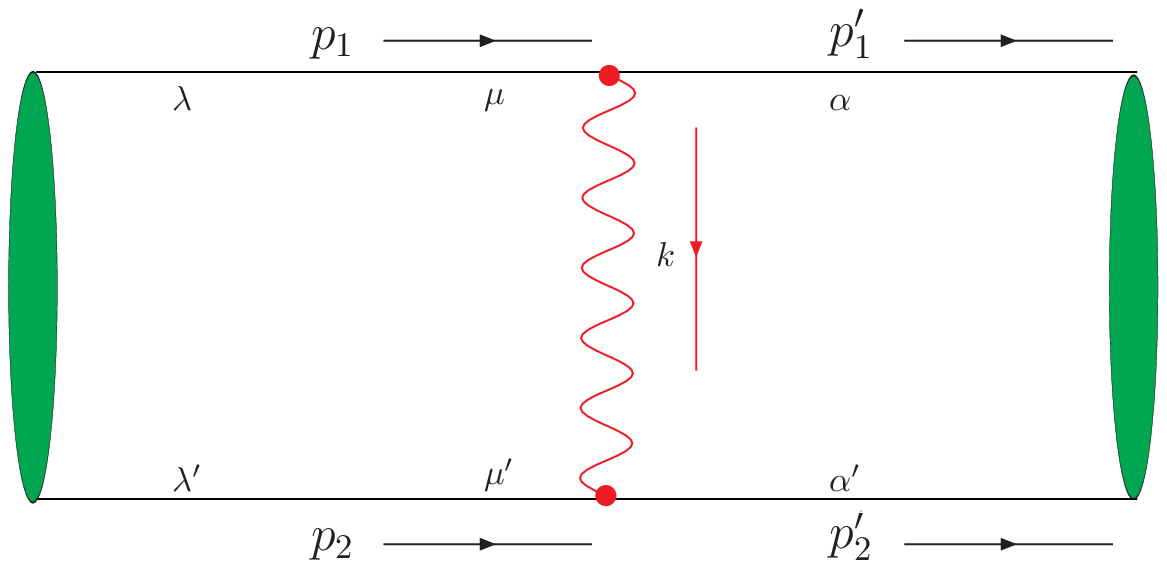}}\,\,\,\,\,\,\,\,\,\,\,\,\,\,\,\,\,\,\,\,\scalebox{0.5}{\includegraphics{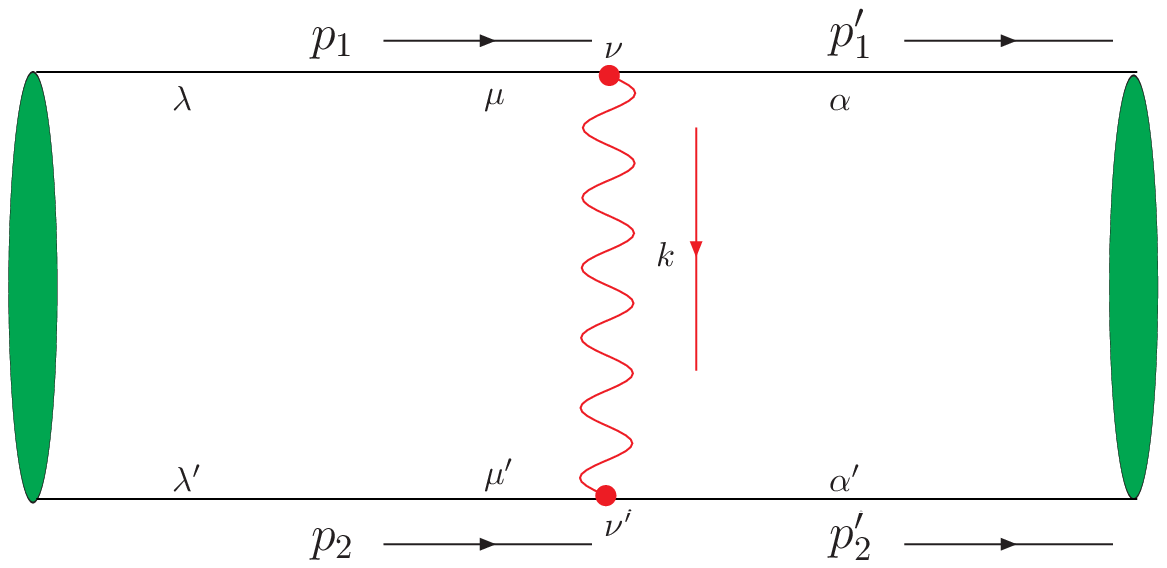}}
\\
(a)\,\,\,\,\,\,\,\,\,\,\,\,\,\,\,\,\,\,\,\,\,\,\,\,\,\,\,\,\,\,\,\,\,\,\,
\,\,\,\,\,\,\,\,\,\,\,\,\,\,\,\,\,\,\,\,\,\,\,\,\,\,\,\,\,\,\,\,\,\,\,\,
\,\,\,\,\,\,\,\,\,\,\,\,\,\,\,\,\,\,\,\,\,\,\,\,\,\,\, (b)
\end{tabular}
\caption{A bound state composed of two vectors. (a) $\pi$ is
exchanged and in fact there totally are four diagrams  like those
in Fig.\ref{DM23}. (b) $\rho$ ($\omega$) is exchanged and there
are four (two) diagrams like those in Fig. \ref{DM23}.
}\label{DM24}
\end{figure}
\end{center}

\subsection{The bound state ($0^+$) composed of two vector mesons}

The quantum number $J^P$ of the bound state composed of two
vectors can be $0^+$, $1^+$ and $2^+$. The corresponding B-S wave
function $|{S'}\rangle$ is  defined as following:
\begin{eqnarray}\label{definition-BS}  \langle 0 | {\rm
T}\,\phi_{1}^\mu(x_1)\phi^\nu_2(x_2) |{S'} \rangle =
{\chi}_{{}_{{0}}}^{}(x_1,x_2)g^{\mu\nu}\,.
\end{eqnarray}
The equation for the B-S wave function is derived as
\begin{eqnarray} \label{4-dim-BS21}
\chi_{_{{}_{0}}}({
p})=\frac{1}{4}\Delta_{1\mu\lambda}\int{d^4{p}'\over(2\pi)^4}\,K_0^{\alpha\alpha'\mu\mu'}({
p},{ p}')\chi_{_{{}_{0}}}^{}({
p}')\Delta_{2\mu'\lambda'}g_{\alpha\alpha'}g^{\lambda\lambda'},
\end{eqnarray}
where
$\Delta_{j\mu\lambda}=\frac{i}{p_j^2-m_j^2}(\frac{p_{j\mu}p_{j\lambda}}{m_2^2}-g_{\mu\lambda})$.

With the Feynman diagrams depicted in Fig. \ref{DM24} and the
effective interaction we obtain
\begin{eqnarray}\label{k0}
  K_{0}^{\alpha\alpha'\mu\mu'}(p,p')&&=K_{01}^{\alpha\alpha'\mu\mu'}(p,p',m_\rho)+K_{02}^{\alpha\alpha'\mu\mu'}(p,p',m_\rho)+
  K_{01}^{\alpha\alpha'\mu\mu'}(p,p',m_\omega)\nonumber\\&&+K_{02}^{\alpha\alpha'\mu\mu'}(p,p',m_\omega)+
  K_{03}^{\alpha\alpha'\mu\mu'}(p,p',m_\pi)+K_{04}^{\alpha\alpha'\mu\mu'}(p,p',m_\pi),\nonumber\\K_{01}^{\alpha\alpha'\mu\mu'}(p,p',m_V)&&=iC_{01}\frac{{q^\nu
q^{\nu'}}/{{m_V}^2}-g^{\nu{\nu'}}}{q^2-m_{V}^2}[g_{_{D^*D^*V}}g^{\alpha\mu}
(p_1+p_1')_\nu-2g_{_{D^*D^*V}}'({q}^{\alpha} g_{\mu\nu}-{q}^\mu
g_{\alpha\nu})]\nonumber\\&&[g_{_{D^*D^*V}}g^{\alpha'\mu'}
(p_2+p_2')_{\nu'}+2g_{_{D^*D^*V}}'({q}^{\alpha'}
g_{\mu'\nu'}-{q}^{\mu'} g_{\alpha'\nu'})]F(q)^2,\nonumber\\
K_{02}^{\alpha\alpha'\mu\mu'}(p,p',m_V)&&=iC_{02}\frac{{q'^\nu
q'^{\nu'}}/{{m_V}^2}-g^{\nu{\nu'}}}{q'^2-m_{V}^2}[g_{_{D^*D^*V}}g^{\alpha\mu}
(p_1+p_2')_\nu-2g_{_{D^*D^*V}}'({q'}^{\alpha} g_{\mu\nu}-{q'}^\mu
g_{\alpha\nu})]\nonumber\\&&[g_{_{D^*D^*V}}g^{\alpha'\mu'}
(p_2+p_1')_{\nu'}+2g_{_{D^*D^*V}}'({q'}^{\alpha'}
g_{\mu'\nu'}-{q'}^{\mu'}
g_{\alpha'\nu'})]F(q')^2,\nonumber\\K_{03}^{\alpha\alpha'\mu\mu'}(p,p',m_P)&&=C_{03}g_{_{D^*D^*P}}^2\varepsilon^{\alpha\beta\mu\nu}q_\nu
(p_1+p_1')_\beta\varepsilon^{\alpha'\beta'\mu'\nu'}q_{\nu'}
(p_2+p_2')_{\beta'}\frac{-i}{q^2-M_\pi^2}F(q)^2,\nonumber\\K_{04}^{\alpha\alpha'\mu\mu'}(p,p',m_P)&&=C_{04}g_{_{D^*D^*P}}^2\varepsilon^{\alpha\beta\mu\nu}q'_\nu
(p_1+p_2')_\beta\varepsilon^{\alpha'\beta'\mu'\nu'}q'_{\nu'}
(p_2+p_1')_{\beta'}\frac{-i}{q'^2-m_P^2}F(q')^2.
\end{eqnarray}
The contributions from vector-exchanges are included in
$K_{01}^{\alpha\alpha'\mu\mu'}(p,p',m_V)$ and
$K_{02}^{\alpha\alpha'\mu\mu'}(p,p',m_V)$ and those for exchanging
pseudoscalars are included in
$K_{03}^{\alpha\alpha'\mu\mu'}(p,p',m_P)$ and
$K_{04}^{\alpha\alpha'\mu\mu'}(p,p',m_P)$. When the bound state is
an isospin-scalar $C_{01}=-\frac{3}{2}$ and  $C_{02}=\frac{3}{2}$
for $\rho$, $C_{01}=\frac{1}{2}$ and $C_{02}=-\frac{1}{2}$ for
$\omega$ and $C_{03}=-\frac{3}{2}$ and $C_{04}=\frac{3}{2}$ for
$\pi$. When the bound state is an isospin-vector
$C_{01}=\frac{1}{2}$ and $C_{02}=\frac{1}{2}$ for $\rho$,
$C_{01}=\frac{1}{2}$ and $C_{02}=\frac{1}{2}$ for $\omega$ and
$C_{03}=\frac{1}{2}$ and $C_{04}=\frac{1}{2}$ for $\pi$.

Defining $ K_{0}(p,p')=\frac{1}{4}{ K_0^{\alpha\alpha'\mu\mu'}}({
p},{ p}')(\frac{p_{2\mu'}
p_{2\lambda'}}{m_2^2}-g_{\mu'\lambda'})(\frac{{p_{1}}_\mu
{p_{1}}_{\lambda}}{m_1^2}-g_{\mu\lambda})$ we derive the B-S
equation which is similar to Eq. (\ref{3-dim-BS1}) but possesses
a different kernel.

The B-S equation can be reduced to
\begin{eqnarray} \label{3-dim-BS3}
{E^2-(E_1+E_2)^2\over (E_1+E_2)/E_1E_2}
\mathcal{\psi}_{{}_0}^{}({\bf p}) ={i\over
2}\int{d^3\mathbf{p}'\over(2\pi)^3}\, {\overline{} K_{{0}}}({\bf
p},{\bf p}')\mathcal{\psi}_{{}_0}^{}({\bf p}')\,,
\end{eqnarray}
where
\begin{eqnarray}
K_{0}(\mathbf{p},\mathbf{p}')&&=K_{01}(\mathbf{p},\mathbf{p}',m_\rho)+{
K_{02}}(\mathbf{p},\mathbf{p}',m_\rho)+{
K_{01}}(\mathbf{p},\mathbf{p}',m_\omega)\nonumber\\&&+{
K_{02}}(\mathbf{p},\mathbf{p}',m_\omega)+{
K_{03}}(\mathbf{p},\mathbf{p}',m_\pi)+{
K_{04}}(\mathbf{p},\mathbf{p}',m_\pi),
\end{eqnarray}
The expressions of $K_{01}(\mathbf{p},\mathbf{p}',m_V)$,
$K_{02}(\mathbf{p},\mathbf{p}',m_V)$, ${
K_{03}}(\mathbf{p},\mathbf{p}',m_P)$ and ${
K_{04}}(\mathbf{p},\mathbf{p}',m_P)$ can be found  in Appendix B.

\subsection{The B-S equation of $1^+$ state which is composed of two vectors}

The B-S wave function of $1^+$ state $|{V'}\rangle$ composed of two axial-vectors is defined as
 \begin{eqnarray} \label{4-dim-BS22}
\langle0|T\phi_\alpha(x_1)\phi_{\alpha'}(x_2)|{V'}\rangle=\frac{\varepsilon_{\alpha\alpha'\tau\tau'}}{\sqrt{6}M}\chi_{_{{1}}}(x_1,x_2)\varepsilon^\tau
P^{\tau'} ,
\end{eqnarray}
where $\varepsilon$ is the polarization vector of $1^+$ state.

The corresponding B-S equation is
\begin{eqnarray} \label{4-dim-BS42}
\chi_{_{{1}}}(p)
=\frac{1}{6M^2}\varepsilon^{\lambda \lambda'
\omega\sigma}\varepsilon_\sigma P_\omega\Delta_{1\mu\lambda}\int{d^4{p'}\over(2\pi)^4}\,K_1^{\alpha\alpha'\mu\mu'}(p,p')
\varepsilon_{\alpha\alpha'\omega'\sigma'}\chi_{_{{1}}}(p')\varepsilon^{\sigma'}P^{\omega'}\Delta_{2\mu'\lambda'}\,,
\end{eqnarray}
where $K_1^{\alpha\alpha'\mu\mu'}(p,p')$ is the same as $K_0^{\alpha\alpha'\mu\mu'}(p,p')$ in Eq.(\ref{k0}).

Defining $ K_{1}(p,p')=\frac{{
K_1^{\alpha\alpha'\mu\mu'}}({ p},{
p}')}{6M^2}\varepsilon^{\lambda \lambda'
\omega\sigma}\varepsilon_\sigma P_\omega(\frac{p_{2\mu'}
p_{2\lambda'}}{m_2^2}-g_{\mu'\lambda'})(\frac{{p_{1}}_\mu
{p_{1}}_{\lambda}}{m_1^2}-g_{\mu\lambda})\varepsilon_{\alpha\alpha'\omega'\sigma'}\varepsilon^{\sigma'}P^{\omega'}$
the B-S equation is reduced to
\begin{eqnarray} \label{3-dim-BS4}
{E^2-(E_1+E_2)^2\over (E_1+E_2)/E_1E_2}
\mathcal{\psi}_{{}_{1}}^{}({\bf p}) ={i\over
2}\int{d^3\mathbf{p}'\over(2\pi)^3}\, {\overline{} K_{1}}({\bf
p},{\bf p}')\mathcal{\psi}_{{}_{1}}^{}({\bf p}')\,,
\end{eqnarray}
with
\begin{eqnarray}
K_{1}(\mathbf{p},\mathbf{p}')&&=K_{11}(\mathbf{p},\mathbf{p}',m_\rho)+{
K_{12}}(\mathbf{p},\mathbf{p}',m_\rho)+{
K_{11}}(\mathbf{p},\mathbf{p}',m_\omega)\nonumber\\&&+{
K_{12}}(\mathbf{p},\mathbf{p}',m_\omega)+{
K_{13}}(\mathbf{p},\mathbf{p}',m_\pi)+{
K_{14}}(\mathbf{p},\mathbf{p}',m_\pi).
\end{eqnarray}
The expressions of $K_{11}(\mathbf{p},\mathbf{p}',m_V)$,
$K_{12}(\mathbf{p},\mathbf{p}',m_V)$, ${
K_{13}}(\mathbf{p},\mathbf{p}',m_P)$ and ${
K_{14}}(\mathbf{p},\mathbf{p}',m_P)$ can be found  in Appendix B.

\subsection{The B-S equation of $2^+$ state $|T'\rangle$ which is composed of two wectors}

 The B-S wave-function of $2^+$ state composed of two axial-vectors is written as
 \begin{eqnarray} \label{4-dim-BS23}
\langle0|T\phi^\alpha(x_1)\phi^{\alpha'}(x_2)|T'\rangle=\frac{1}{\sqrt{5}}\chi_{_{2}}(x_1,x_2)\varepsilon^{\alpha\alpha'}
,
\end{eqnarray}
where $\varepsilon$ is the polarization vector of the $2^+$ state.

The B-S equation can be expressed as
\begin{eqnarray} \label{4-dim-BS43}
\chi_{_{2}}(p)
=\frac{1}{5}\varepsilon^{\lambda\lambda'}\Delta_{1\mu\lambda}\int{d^4{q}\over(2\pi)^4}\,K_2^{\alpha\alpha'\mu\mu'}(p,p')
\varepsilon_{\alpha\alpha'}\chi_{_{2}}(q)\Delta_{2\mu'\lambda'}\,,
\end{eqnarray}
where $K_2^{\alpha\alpha'\mu\mu'}(p,p')$ is the same as $K_0^{\alpha\alpha'\mu\mu'}(p,p')$ in Eq.(\ref{k0}).

Defining $K_{2}(p,p')= \frac{{
K_2^{\alpha\alpha'\mu\mu'}}({ p},{
p}')}{5}\varepsilon^{\lambda \lambda'
}(\frac{p_{2\mu'}
p_{2\lambda'}}{m_2^2}-g_{\mu'\lambda'})(\frac{{p_{1}}_\mu
{p_{1}}_{\lambda}}{m_1^2}-g_{\mu\lambda})\varepsilon_{\alpha\alpha'}$
the B-S equation can be reduced to
\begin{eqnarray} \label{3-dim-BS5}
{E^2-(E_1+E_2)^2\over (E_1+E_2)/E_1E_2}
\mathcal{\psi}_{{}_{2}}^{}({\bf p}) ={i\over
2}\int{d^3\mathbf{p}'\over(2\pi)^3}\, {\overline{} K_{2}}({\bf
p},{\bf p}')\mathcal{\psi}_{{}_{2}}^{}({\bf p}')\,,
\end{eqnarray}
where
\begin{eqnarray}
K_{2}(\mathbf{p},\mathbf{p}')&&=K_{21}(\mathbf{p},\mathbf{p}',m_\rho)+{
K_{22}}(\mathbf{p},\mathbf{p}',m_\rho)+{
K_{21}}(\mathbf{p},\mathbf{p}',m_\omega)\nonumber\\&&+{
K_{22}}(\mathbf{p},\mathbf{p}',m_\omega)+{
K_{23}}(\mathbf{p},\mathbf{p}',m_\pi)+{
K_{24}}(\mathbf{p},\mathbf{p}',m_\pi).
\end{eqnarray}
The expressions of $K_{21}(\mathbf{p},\mathbf{p}',m_V)$,
$K_{22}(\mathbf{p},\mathbf{p}',m_V)$, ${
K_{23}}(\mathbf{p},\mathbf{p}',m_P)$ and ${
K_{24}}(\mathbf{p},\mathbf{p}',m_P)$ can be found  in Appendix B.

\section{Numerical results}
Now let us solve the B-S equations  (\ref{3-dim-BS1}),
(\ref{3-dim-BS2}), (\ref{3-dim-BS3}), (\ref{3-dim-BS4}) and
(\ref{3-dim-BS5}). Since we are interested in the ground state of a bound state the function $\psi_{J}(\mathbf{p})$ ($J$ represents $S, V, 0, 1$ or
$2$) only depends on
the norm of the three-momentum and we may first integrate over the
azimuthal angle of the functions in (\ref{3-dim-BS1}),
(\ref{3-dim-BS2}), (\ref{3-dim-BS3}), (\ref{3-dim-BS4}) or
(\ref{3-dim-BS5})
$$\frac{i}{2}\int{d^3\mathbf{p}'\over(2\pi)^3}\, {\overline{} K_J}({\bf p},{\bf
p}'),  $$  to obtain a potential form
$U_J(|\mathbf{p}|,|\mathbf{p}'|)$ , then the B-S equation turns into a one-dimension integral
equation
\begin{eqnarray} \label{3-dim-BS6}
\psi_J({\bf |p|}) ={(E_1+E_2)/E_1E_2\over E^2-(E_1+E_2)^2 }\int{d
\mathbf{|p}'|}\, {\overline{} U_J}({\bf |p|},{\bf
|p}'|)\psi_J({\bf |p}'|) .
\end{eqnarray}
When the potential
$U_{J}(\mathbf{p},\mathbf{p}')$  is attractive and strong enough
the corresponding B-S equation has a solution(s) and we can obtain
the mass of the possible bound state.

Generally the standard way of solving an integral equation is to
discretize and perform algebraic operations. Concretely,  we let
$\bf |p|$ and $\bf |p'|$ take $n$ ( $n$ is sufficiently large)
order discrete values $Q_1$, $Q_2$,...$Q_n$ and the gap between
two adjacent values be $\Delta  Q$, then the integral equation is
transformed into $n$ coupled algebraic equations.
$\psi_{{}_J}^{}(Q_1),\psi_{{}_J}^{}(Q_2),...\psi_{{}_J}^{}(Q_n)$ (
the subscript $J$ denotes $S$, $V$, 0, 1 or 2) constitute a
column matrix and the coefficients would stand as an $n\times n$
matrix $M$, thus these algebraic equations can be regarded as a matrix
equation with a unique eigenvalue  1. If one can obtain a value of $E$
which satisfies the equation with reasonable input parameters
and $E$ is not far from $E_1+E_2$ the corresponding eigenvector should exist as a bound state.

In our calculation the values of the parameters $g_{_{DDV}},
g_{_{DD^*P}}, g_{_{DD^*V}}$, $g_{_{D^*D^*V}}$ and
$g'_{_{D^*D^*V}}$ are presented in Appendix A.
 Supposing $T^+_{cc}$ is a $D^0D^{*+}$ bound state, by fitting its mass  we fix
 $\Lambda=1.134$ GeV. In Ref. \cite{Cheng:2004ru,Meng:2007tk} the authors
suggested a relation: $\Lambda=m+\alpha \Lambda_{QCD}$ where $m$
is the mass of the exchanged meson, $\alpha$ is a number of $O(1)$
and $\Lambda_{QCD}=220$ MeV i.e. $\Lambda\sim 1$GeV for exchanging
$\rho$ or $\omega$. The value of $\Lambda$ we obtained locates within the
range.

The masses of the concerned constituent mesons $m_D$ , $m_{D^*}$
, $m_B$ and $m_{B^*}$ are directly taken from the databook
\cite{PDG10}.

\subsection{The results of $D^{(*)}D^{(*)}$ system}

Now let us try to calculate the eigenvalues of these  systems of $D^0
D^+(J^P=0^+, I=1)$, $D^0 D^{*+}(J^P=1^+, I=1)$, $D^0
D^{*+}(J^P=1^+, I=0)$, $D^{*0}  D^{*+}(J^P=0^+, I=1)$, $D^{*0}
D^{*+}(J^P=1^+, I=0)$ and $D^{*0}  D^{*+}(J^P=2^+, I=1)$
respectively. Apparently with the parameters $\Lambda$ and
coupling constants, not all B-S equations are solvable. For
$D^0D^{*+}$ system with $I=0$ the B-S equation has a solution. It
implies that $D^0D^{*+}$ can form an isospin scalar bound state by
exchanging light mesons. In Ref.\cite{Zhao:2021cvg} the authors also obtained the same results with similar approach.
For the $D^0D^{*+}$ ($I=1$) or $D^0D^{+}$ ($I=1$) system, employing a larger
$\Lambda$ and coupling constant we can obtain a solution. It may imply the effective
interaction between the two constituents is weak. For the $D^{*0}D^{*+}$ system  we
can obtain an eigenvalue 18.51 MeV, the corresponding eigenstate is a bound state of
$J=0$ and  $I=1$. In table \ref{tab:ev1} there are many
places symbolized by $``*"$ or $``-"$ which means such bound
states cannot exist due to the symmetry restriction or the B-S
equation has no solution. However in Ref.\cite{Chen:2021cfl} $D^{*}D^{*}$ system with $J=1$ and isospin
$I=0$ was suggested to exist, which contradicts to our result. The reason is that the authors of Ref.\cite{Chen:2021cfl}
did  not
symmetrize and antisymmetrize the flavor wave functions of $D^*D^*$
for
$I=0$ and $I=1$ states\cite{Li:2012ss,Liu:2019stu}. Instead, we redo the calculation as
the total symmetry of the wave-function including
flavor, spin parts and orbital angular momentum is taken into account.
For the $D^{*0}D^{*+}$ system the spin wave-function is symmetrized and/or
antisymmetrized so that the flavor wave functions need to be correspondingly symmetrized and
antisymmetrized when $l=0$. For $I=1$ and $I=0$ states of $D^{*0}D^{*+}$ the symmetric and
antisymmetric flavor wave-functions were considered in Ref.\cite{Deng:2021gnb}.

\subsection{The results of the $B^{(*)}B^{(*)}$ system}

Considering the flavor $SU(3)$ symmetry and heavy quark effective
symmetry we generalize those relations as $g_{_{BBV}}=g_{_{DDV}},
g_{_{BB^*P}}=g_{_{DD^*P}}, g_{_{BB^*V}}= g_{_{DD^*V}}$,
$g_{_{B^*B^*V}}=g_{_{D^*D^*V}}$ and
$g'_{_{B^*B^*V}}=g'_{_{D^*D^*V}}$ which should be a not-bad
approximation.

\begin{table}
\caption{ The binding energy of the ground $DD^{(*)}$ system (in
unit of MeV).} \label{tab:ev1}
\begin{tabular}{c|c|c|c|c|c}\hline\hline
 ~~~~~~~~   &  ~~~$D^0 D^{+}(0^+)$~~~   &
 ~~~$D^0 D^{*+}(1^+)$~~~ &  ~~~$D^{*0} D^{*+}(0^+)$~~~&  ~~~$D^{*0} D^{*+}(1^+)$~~~ &  ~~~$D^{*0} D^{*+}(2^+)$~~~   \\\hline
 $I=0$    &$\times$   & 0.273     & $\times$  &  - &$\times$ \\
 $I=1$    & -   & -     & 18.51   & $\times$ &- \\
\hline\hline
\end{tabular}
\end{table}

\begin{table}
\caption{ The eigenvalues of the ground $BB^{(*)}$ system (in unit
of MeV).} \label{tab:ev2}
\begin{tabular}{c|c|c|c|c|c}\hline\hline
  ~~~~~~~~   &  ~~~$B^0 B^{+}(0^+)$~~~   &
 ~~~$B^0 B^{*+}(1^+)$~~~ &  ~~~$B^{*0} B^{*+}(0^+)$~~~&  ~~~$B^{*0} B^{*+}(1^+)$~~~ &  ~~~$B^{*0} B^{*+}(2^+)$~~~   \\\hline
 $I=0$    &$\times$   & 17.38     & $\times$   &  - &$\times$ \\
 $I=1$    & -   & -     &  62.37   & $\times$ &- \\
\hline\hline
\end{tabular}
\end{table}

We use the same parameter $\Lambda$ fixed for the $DD^*$ systems
to solve the B-S equation for the $B^{(*)}B^{(*)}$ systems. We
find taht two states which are the counterparts of $D^{(*)}D^{(*)}$ can
exist. The binding energy of each state shown in table
\ref{tab:ev2} is apparently larger than that of the corresponding
state of $D^{(*)}D^{(*)}$ since the mass of $B^{(*)}$ meson is
larger than that of $D^{(*)}$ meson.

\begin{center}
\begin{figure}[htb]
\begin{tabular}{cc}
\scalebox{0.7}{\includegraphics{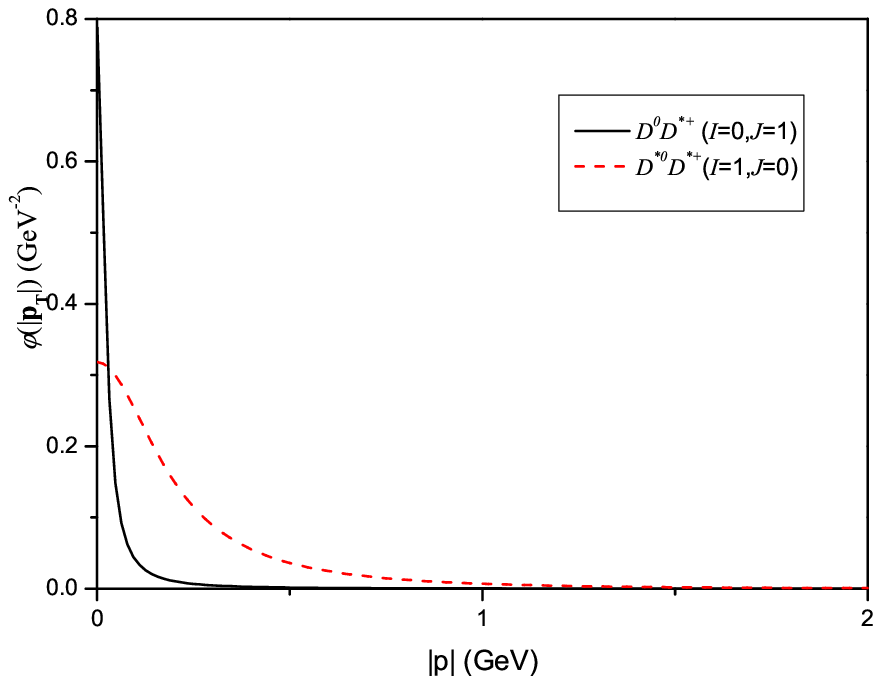}}\,\,\,\,\,
\,\,\,\,\,\,\,\scalebox{0.7}{\includegraphics{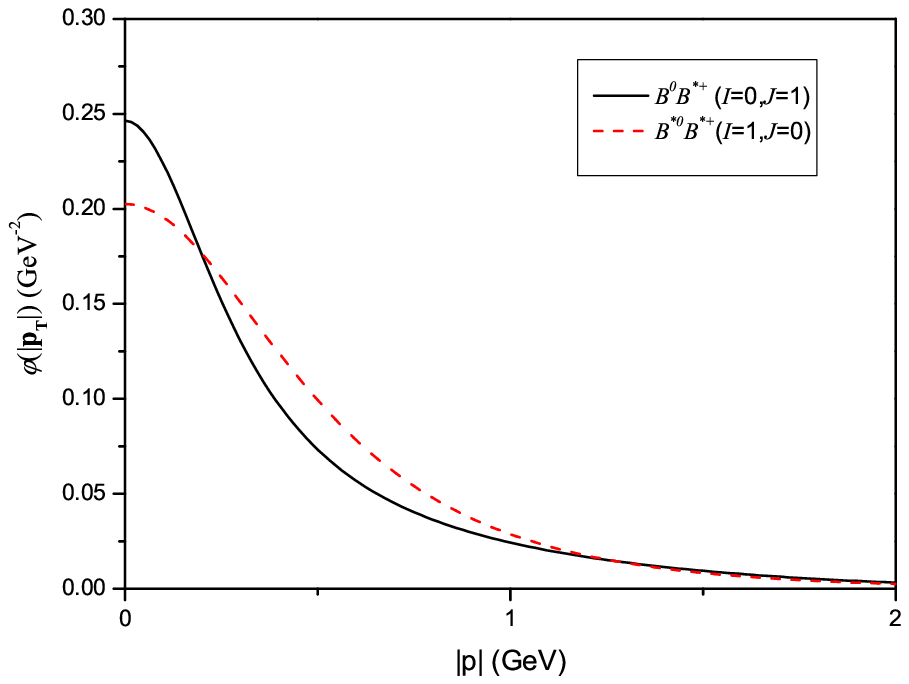}}\\
(a)\,\,\,\,\,\,\,\,\,\,\,\,\,\,\,\,\,\
,\,\,\,\,\,\,\,\,\,\,\,\,\,\,\,\,\,\,
\,\,\,\,\,\,\,\,\,\,\,\,\,\,\,\,\,\,\,\,\,\,\,\,\,\,\,\,\,\,\,\,\,\,\,\,
\,\,\,\,\,\,\,\,\,\,\,\,\,\,\,\,\,\,\,\,\,\,\,\,\,\,\,\,\,\,\,\,\,\,\,\,
\,\,\,\,\,\,\,\,\,\,\,\,\,\,\, (b)
\end{tabular}
\caption{ the unnormalized wave functions of the bound
states.}\label{DM25}
\end{figure}
\end{center}

\section{A brief summary}
In this work we study whether two charmed (or bottomed) mesons can
form a hadronic molecule. We employ the B-S framework to search
for possible bound states of $D^{(*)}D^{(*)}$\cite{Dai:2021vgf} and
$B^{(*)}B^{(*)}$. In Ref.
\cite{Guo:2007mm,Feng:2011zzb,Ke:2012gm,Ke:2020eba,Ke:2021iyh} the
B-S wave functions for the systems of  one vector and one
pseudoscalar, two pseudoscalar mesons and two vectors were
studied. It is noted that all those works are dealing with bound
states made of one particle and one-antiparticle, no matter they
are pseudoscalar or vector bosons. In comparison, this work is
concerning particle-particle bound states(charmed $D^{(*)}D^{(*)}$
or bottomed $B^{(*)}B^{(*)}$). Since the two constituents are
accounted as identical, symmetrization of the total wavefunctions
is necessarily required. In this work we deduce the interaction
kernels for these systems and solve these B-S equations.

In order to obtain the interaction kernels for B-S equations we use
the heavy meson chiral perturbation theory to calculate the
corresponding Feynman diagrams where $\pi$, $\rho$ or $\omega$ are
exchanged. All coupling constants are taken from relevant references. For
making predictions we use the binding energy of $T_{cc}^+$
to fix the parameter $\Lambda$ under the hypothesis that
$T_{cc}^+$ is a bound state of $D^0D^{*+}$ with $I=0$ and $J=1$.
With the same parameters we confirm that $D^{*0}D^{*+}$ with $I=1$ and
$J=0$ should exist. For the $D^{*0}D^{*+}$ system with $I=1$ and $J=2$, a larger $\Lambda$ or
large coupling constants are needed to form bound states.

Considering the flavor $SU(3)$ symmetry and heavy quark spin
symmetry we employ the same parameters to calculate possible bound
states of $B^{(*)}B^{(*)}$. Two states which are the counterparts
of $D^{(*)}D^{(*)}$ can exist. The binding energy of each state is
apparently larger than that of the corresponding state of
$D^{(*)}D^{(*)}$ since  $B^{(*)}$ meson is heavier than $D^{(*)}$
meson.

Since the parameters are fixed from data which span
a relatively large range we cannot expect all the numerical
results to be very accurate. The
goal of this work is to study whether two charmed (or
bottomed) mesons can form a molecular state. Our results, even if not
accurate, have obvious qualitative significance. Definitely,
further theoretical and experimental works are badly needed for
gaining better understanding of these exotic hadrons.

\section*{Acknowledgments}

 This work is supported by the National Natural Science Foundation
of China (NNSFC) under the contract No. 12075167, 11975165, 11735010, 12035009 and 12075125. We thank Prof. Xiang Liu for his valuable discussion.
\appendix

\section{The effective interactions}
The effective interactions can be found
in\cite{Colangelo:2005gb,Colangelo:2012xi,Ding:2008gr}
\begin{eqnarray}
&&\mathcal{L}_{_{DDV}}=g_{_{DDV}}(D_{b}\stackrel{\leftrightarrow}{\partial}_{\beta}
D^{\dag}_{a})(
\mathcal{V}^\beta)_{ba},\\&&\mathcal{L}_{_{DD^*V}}=ig_{_{DD^*V}}\varepsilon^{\alpha\beta\mu\nu}({\partial}_{\alpha}\mathcal{V}_\beta
-{\partial}_{\beta}\mathcal{V}_\alpha)_{ba}(\partial_\nu
D_{b}D^{*\mu\dag}_{a}-\partial_\nu
D^{*\mu\dag}_{b}D_{a}),\\&&\mathcal{L}_{_{DD^*P}}=g_{_{DD^*P}}D_{b}(\partial_\mu
\mathcal{M})_{ba}D^{*\mu\dag}_{a}+g_{_{DD^*P}}D^{*\mu}_{b}(\partial_\mu
\mathcal{M})_{ba}D^{\dag}_{a},\\&&\mathcal{L}_{_{D^*D^*P}}=g_{_{D^*D^*P}}(D^{*\mu}_{b}\stackrel{\leftrightarrow}{\partial}^{\beta}
D^{*\alpha\dag}_{a})(\partial^\nu
\mathcal{M})_{ba}\varepsilon_{\nu\mu\alpha\beta},
\\&&\mathcal{L}_{_{D^*D^*V}}=ig_{_{D^*D^*V}}(D^{*\nu}_{b}\stackrel{\leftrightarrow}{\partial}_{\mu}
D^{*\dag}_{a\nu})(
\mathcal{V})_{ba}^\mu+ig'_{_{D^*D^*V}}(D^{*\mu}_{b}
D^{*\nu\dag}_{a}-D^{*\mu\dag}_{b} D^{*\nu}_{a})(
\partial_\mu\mathcal{V}_\nu-\partial_\nu\mathcal{V}_\mu)_{ba}
\end{eqnarray}
where $a$ and $b$ represent the flavors of light quarks. In
Refs.\cite{Ding:2008gr} $\mathcal{M}$ and $\mathcal{V}$ are
$3\times 3$ hermitian and traceless matrixs $
\left(\begin{array}{ccc}
        \frac{\pi^0}{\sqrt{2}}+\frac{\eta}{\sqrt{6}} &\pi^+ &K^+ \\
         \pi^- & -\frac{\pi^0}{\sqrt{2}}+\frac{\eta}{\sqrt{6}}&K^0\\
         K^-& \bar{K^0} & -\sqrt{\frac{2}{3}}\eta
      \end{array}\right)$
       and $
\left(\begin{array}{ccc}
        \frac{\rho^0}{\sqrt{2}}+\frac{\omega}{\sqrt{2}} &\rho^+ &K^{*+} \\
         \rho^- & -\frac{\rho^0}{\sqrt{2}}+\frac{\omega}{\sqrt{2}}&K^{*0}\\
         K^{*-}& \bar{K^{*0}} & \phi
      \end{array}\right)$ respectively.

In the chiral and heavy quark limit, the above coupling constants
are $g_{_{DDV}}=\frac{\beta g_V}{\sqrt{2}},
g_{_{DD^*V}}=\frac{\lambda g_V}{\sqrt{2}},
g_{_{D^*D^*P}}=\frac{g}{f_\pi},$ $g_{_{DD^*P}}=-\frac{2g}{f_{\pi}}
\sqrt{M_{D}M_{D^*}}, g_{_{D^*D^*V}}=-\frac{\beta
g_V}{\sqrt{2}},\,\, g'_{_{D^*D^*V}}=-\sqrt{2}\lambda g_V M_{D^*}$
with $f_\pi=132$ MeV\cite{Colangelo:2005gb},
$g=0.64$\cite{Colangelo:2012xi}, $\kappa=g$, $\beta=0.9$,
$g_V=5.9$\cite{Falk:1992cx} and $\lambda =0.56$ GeV$^{-1}$\cite{Chen:2019asm}.
\section{kernel}

\begin{eqnarray}
&&K_{S0}({\bf p},{\bf p}',m_v) = i C_{S1}\,g_{_{DDV}}^2\, {({\bf
p}+{\bf p}')^2 + 4\eta_1\eta_2 E^2 + ({\bf p}^2-{\bf
p}'{}^2)^2/m_{ V}^2 \over ({\bf p}-{\bf p}')^2 + m_{ V}^2}F(\bf
q)^2. \label{potential-with-isospin-factor}
\end{eqnarray}

\begin{eqnarray}
{K_{V1}}(\mathbf{p},\mathbf{p}',m_{ V})
&&=\frac{iC_{V1}g_{_{DDV}}F({\bf q})^2}{-(\mathbf{p} -
\mathbf{p}')^2-m_{V}^2}\{-g_{_{D^*D^*V}} \mathbf{p}^2 [4 \eta_1
\eta_2 M^2 + (\mathbf{p} + \mathbf{p}')^2]/m_2^2\nonumber\\&&-3
g_{_{D^*D^*V}} (\mathbf{p}^2 -
\mathbf{p}'^2)^2/m_V^2-g_{_{D^*D^*V}} \mathbf{p}^2(\mathbf{p}^2 -
\mathbf{p}'^2)^2/(m_2^2 m_V^2)\nonumber\\&&+4 \eta_1 \eta_2
g'_{_{D^*D^*V}} M^2 (-\mathbf{p}^2 + \mathbf{p}\cdot
\mathbf{p}')/m_2^2\}\nonumber\\{K_{V2}}(\mathbf{p},\mathbf{p}',m_{
V})&& = \frac{i C_{V2}g^2_{_{D^*DV}}F(\bf
q')^2}{-(\mathbf{p}+\mathbf{p'})^2-m_{V}^2}[-8 \eta_1\eta_2 M^2
(\mathbf{p}-\mathbf{p}')^2-4 \mathbf{p}^2 \mathbf{p}'^2 + 4
\mathbf{p}\cdot
\mathbf{p}'^2]\nonumber\\K_{V3}(\mathbf{p},\mathbf{p}',m_P)&&=\frac{i
C_{V3}g_{_{DD^*P}}^2F(\bf
q')^2}{-(\mathbf{p}+\mathbf{p'})^2-m_{P}^2}[-(\mathbf{p} +
\mathbf{q})^2-(\mathbf{p}^2 (\mathbf{p} + \mathbf{q})^2/m2^2)]
\end{eqnarray}

\begin{eqnarray}
K_{01}&&=\frac{iC_{01} g^2_{_{D^*D^*V}} F(\bf q)^2}{4[-(\mathbf{p}
- \mathbf{p}')^2-m_{V}^2]}\{-16 \eta_1 \eta_2 M^2-4 ({\bf p}+{\bf
p}')^2-\frac{(\eta_1 \eta_2 M^2+{\bf p}^2)^2 ({\bf p}^2-{\bf
p}'^2)^2}{(m_1^2 m_2^2 m_V^2)}\nonumber\\&&+(\eta_1^2 M^2-{\bf
p}^2) ({\bf p}^2-{\bf p}'^2)^2/(m_1^2 m_V^2)+(\eta_2^2 M^2-{\bf
p}^2) ({\bf p}^2-{\bf p}'^2)^2/(m_2^2 m_V^2)\nonumber\\&&-4 ({\bf
p}^2-{\bf p}'^2)^2/m_V^2 -(\eta_1 \eta_2 M^2+{\bf p}^2)^2 [4
\eta_1 \eta_2 M^2+({\bf p}+{\bf p}')^2]/(m_1^2
m_2^2)\nonumber\\&&+(\eta_1^2 M^2-{\bf p}^2) [4 \eta_1 \eta_2
M^2+({\bf p}+{\bf p}')^2]/m_1^2+(\eta_2^2 M^2-{\bf p}^2) [4 \eta_1
\eta_2 M^2+({\bf p}+{\bf p}')^2]/m_2^2\}\nonumber\\&&
+\frac{iC_{01} g_{_{D^*D^*V}}g'_{_{D^*D^*V}}F(\bf
q)^2}{-(\mathbf{p} - \mathbf{p}')^2-m_{V}^2} (\eta_1+\eta_2)^2 M^2
(\eta_1 \eta_2 M^2+{\bf p}^2) ({\bf p}^2-{\bf p}\cdot{\bf p}')
/(m_1^2 m_2^2)\nonumber\\&& +\frac{iC_{01}{g'}^2_{_{D^*D^*V}}
F(\bf q)^2}{4[-(\mathbf{p} -  \mathbf{p}')^2-m_{V}^2]} \{[8 ({\bf
p}^2+\eta_1 \eta_2 M^2) ({\bf p}^2 {\bf p}'^2-{\bf p}\cdot{\bf
p}'^2)\nonumber\\&&-4 (\eta_1^2 \eta_2^2 M^4+4 \eta_1 \eta_2 {\bf
p}^2 M^2+3 {\bf p}^4) ({\bf p}-{\bf p}')^2]/(m_1^2
m_2^2)\nonumber\\&&-24 ({\bf p}-{\bf p}')^2+[4 (\eta_1^2 M^2-3
{\bf p}^2) ({\bf p}-{\bf p}')^2+8 ({\bf p}^2 {\bf p}'^2-{\bf
p}\cdot{\bf p}'^2)]/m_1^2\nonumber\\&&+[4 (\eta_2^2 M^2-3 {\bf
p}^2) ({\bf p}-{\bf p}')^2+8 ({\bf p}^2 {\bf p}'^2-{\bf
p}\cdot{\bf p}'^2)]/m_2^2\} \nonumber\\K_{02}=&&\frac{iC_{02}
g^2_{_{D^*D^*V}} F(\bf q')^2}{4[-(\mathbf{p} +
\mathbf{p}')^2-m_{V}^2]}\{-16 \eta_1 \eta_2 M^2-4 ({\bf p}-{\bf
p}')^2-\frac{(\eta_1 \eta_2 M^2+{\bf p}^2)^2 ({\bf p}^2-{\bf
p}'^2)^2}{(m_1^2 m_2^2 m_V^2)} \nonumber\\&&+(\eta_1^2 M^2-{\bf
p}^2) ({\bf p}^2-{\bf p}'^2)^2/(m_1^2 m_V^2)+(\eta_2^2 M^2-{\bf
p}^2) ({\bf p}^2-{\bf p}'^2)^2/(m_2^2 m_V^2) \nonumber\\&&-(4
({\bf p}^2-{\bf p}'^2)^2)/m_V^2- (\eta_1 \eta_2 M^2+{\bf p}^2)^2
[4 \eta_1 \eta_2 M^2+({\bf p}-{\bf p}')^2]/(m_1^2
m_2^2)\nonumber\\&&+(\eta_1^2 M^2-{\bf p}^2) [4 \eta_1 \eta_2
M^2+({\bf p}-{\bf p}')^2]/m_1^2+(\eta_2^2 M^2-{\bf p}^2) [4 \eta_1
\eta_2 M^2+({\bf p}-{\bf p}')^2]/m_2^2\} \nonumber\\&&
+\frac{iC_{02} g_{_{D^*D^*V}}g'_{_{D^*D^*V}} F(\bf
q')^2}{-(\mathbf{p} + \mathbf{p}')^2-m_{V}^2} (\eta_1+\eta_2)^2
M^2 (\eta_1 \eta_2 M^2+{\bf p}^2) ({\bf p}^2+{\bf p}\cdot{\bf p}')
/(m_1^2 m_2^2)\nonumber\\&& +\frac{iC_{02} {g'}^2_{_{D^*D^*V}}
F(\bf q')^2}{4[-(\mathbf{p} + \mathbf{p}')^2-m_{V}^2]} \{[8 ({\bf
p}^2+\eta_1 \eta_2 M^2) ({\bf p}^2 {\bf p}'^2-{\bf p}\cdot{\bf
p}'^2)\nonumber\\&&-4 (\eta_1^2 \eta_2^2 M^4+4 \eta_1 \eta_2 {\bf
p}^2 M^2+3 {\bf p}^4) ({\bf p}+{\bf p}')^2]/(m_1^2
m_2^2)\nonumber\\&&-24 ({\bf p}+{\bf p}')^2+[4 (\eta_1^2 M^2-3
{\bf p}^2) ({\bf p}+{\bf p}')^2+8 ({\bf p}^2 {\bf p}'^2-{\bf
p}\cdot{\bf p}'^2)]/m_1^2\nonumber\\&&+[4 (\eta_2^2 M^2-3 {\bf
p}^2) ({\bf p}+{\bf p}')^2+8 ({\bf p}^2 {\bf p}'^2-{\bf
p}\cdot{\bf p}'^2)]/m_2^2\}
\nonumber\\
K_{03}&&=2 \frac{iC_{03} g^2_{_{D^*D^*P}} F(\bf
q)^2}{(\mathbf{p} - \mathbf{p}')^2+m_{P}^2} [-{\bf p}\cdot{\bf p}'^2+\eta_1 \eta_2 M^2 ({\bf p}-{\bf p}')^2+{\bf p}^2 {\bf p}'^2]\nonumber\\
K_{04}&&=2\frac{iC_{04} g^2_{_{D^*D^*P}} F(\bf q')^2}{(\mathbf{p}
+ \mathbf{p}')^2+m_{P}^2} [-{\bf p}\cdot{\bf p}'^2 + \eta_1 \eta_2
M^2 ({\bf p} + {\bf p}')^2+ {\bf p}^2 {\bf p}'^2 ]
\end{eqnarray}

\begin{eqnarray}
K_{11}&&=-\frac{iC_{11} g^2_{_{D^*D^*V}} F(\bf q)^2}{-(\mathbf{p}
- \mathbf{p}')^2-m_{V}^2}\{4 \eta_1 \eta_2 M^2 m_V^2+[m_V^2+({\bf
p}-{\bf p}')^2] ({\bf p}+{\bf p}')^2\}\nonumber\\&&[(3 m_2^2+{\bf
p}^2) m_1^2+m_2^2 {\bf p}^2] /(3 m_1^2 m_2^2 m_V^2)\nonumber\\&&
+4\frac{iC_{11} g_{_{D^*D^*V}} g'_{_{D^*D^*V}} F(\bf
q)^2}{-(\mathbf{p} -  \mathbf{p}')^2-m_{V}^2}\eta_1 \eta_2
 M^2 (m_1^2+m_2^2) ({\bf p}^2-{\bf p}\cdot{\bf p}') /(3 m_1^2
m_2^2)\nonumber\\&& - \frac{iC_{11} {g'}^2_{_{D^*D^*V}} F(\bf
q)^2}{-(\mathbf{p} -  \mathbf{p}')^2-m_{V}^2} \{[4(2 m_2^2+{\bf
p}^2) m_1^2+4m_2^2 {\bf p}^2] ({\bf p}-{\bf p}')^2\nonumber\\&&+2
(m_1^2+m_2^2) ({\bf p}\cdot{\bf p}'^2-{\bf p}^2 {\bf p}'^2)\}/(3
m_1^2 m_2^2) \nonumber\\K_{12}&&=-\frac{iC_{12} g^2_{_{D^*D^*V}}
F(\bf q')^2}{-(\mathbf{p} + \mathbf{p}')^2-m_{V}^2}[(3 m_2^2+{\bf
p}^2) m_1^2+m_2^2 {\bf p}^2] \{4 \eta_1 \eta_2 M^2
m_V^2\nonumber\\&&+ [m_V^2+({\bf p}+{\bf p}')^2]({\bf p}-{\bf
p}')^2\}/(3 m_1^2 m_2^2 m_V^2)\nonumber\\&& +4\frac{iC_{12}
g_{_{D^*D^*V}}g'_{_{D^*D^*V}} F(\bf q')^2}{-(\mathbf{p} +
\mathbf{p}')^2-m_{V}^2} \eta_1 \eta_2 M^2 (m_1^2+m_2^2) ({\bf
p}^2+{\bf p}\cdot{\bf p}') /(3 m_1^2 m_2^2) \nonumber\\&&
-\frac{iC_{12} {g'}^2_{_{D^*D^*V}} F(\bf q')^2}{-(\mathbf{p} +
\mathbf{p}')^2-m_{V}^2} \{[4(2 m_2^2+{\bf p}^2) m_1^2+4m_2^2 {\bf
p}^2] ({\bf p}+{\bf p}')^2\nonumber\\&&+2 (m_1^2+m_2^2) ({\bf
p}\cdot{\bf p}'^2-{\bf p}^2 {\bf p}'^2)\}/(3 m_1^2 m_2^2)
\nonumber\\
K_{13}&&=\frac{4}{3}\frac{iC_{13} g^2_{_{D^*D^*P}} F(\bf
q)^2}{-(\mathbf{p} - \mathbf{p}')^2-m_{P}^2}\eta_1 \eta_2 M^2 ({\bf p}-{\bf p}')^2\nonumber\\
K_{14}&&=\frac{4}{3}\frac{iC_{14} g^2_{_{D^*D^*P}} F(\bf
q')^2}{-(\mathbf{p} + \mathbf{p}')^2-m_{P}^2}\eta_1 \eta_2 M^2
({\bf p}+{\bf p}')^2
\end{eqnarray}

\begin{eqnarray}
K_{21}&&=- \frac{iC_{21} g^2_{_{D^*D^*V}} F(\bf q)^2}{-(\mathbf{p}
- \mathbf{p}')^2-m_{V}^2}g^2_{D^*D^*V} [5 m_2^2 {\bf p}^2 + 2 {\bf
p}^4 +   5 m_1^2 (3 m_2^2 +   {\bf p}^2)] \{4 \eta_1 \eta_2 M^2
m_V^2 \nonumber\\&&+ [m_V^2 + ({\bf p} - {\bf
 p}')^2]
  ({\bf p} +  {\bf p}')^2\}/(15 m_1^2 m_2^2 m_V^2)\nonumber\\&&+4 \frac{iC_{21} g_{_{D^*D^*V}}{g'}_{_{D^*D^*V}} F(\bf
q)^2}{-(\mathbf{p} -  \mathbf{p}')^2-m_{V}^2}\eta_1 \eta_2  M^2 (5 m_1^2 + 5 m_2^2 + 4 {\bf
p}^2) ({\bf p}^2 - {\bf p}\cdot{\bf p}')/( 15 m_1^2
m_2^2)\nonumber\\&&
 +
\frac{iC_{21} {g'}^2_{_{D^*D^*V}} F(\bf q)^2}{-(\mathbf{p} -
\mathbf{p}')^2-m_{V}^2} \{ [20 m_1^2 (2 m_2^2 + {\bf p}^2) + 4{\bf
p}^2 (-2 \eta_1 \eta_2 M^2 + 5 m_2^2 + 2 {\bf p}^2)] ({\bf p} -
{\bf p}')^2/( 15 m_1^2 m_2^2)\nonumber\\&& +
    2 (-6 \eta_1 \eta_2 M^2 + 5 m_1^2 + 5 m_2^2 + 2 {\bf p}^2) ({\bf p}\cdot{\bf p}'^2 -
       {\bf p}^2 {\bf p}'^2)/(15 m_1^2 m_2^2)\}
\nonumber\\K_{22}&&=- \frac{iC_{22} g^2_{_{D^*D^*V}} F(\bf
q')^2}{-(\mathbf{p} + \mathbf{p}')^2-m_{V}^2}[5 m_2^2 {\bf p}^2 +
2 {\bf p}^4 +
    5 m_1^2 (3 m_2^2 + {\bf p}^2)] \{4 \eta_1 \eta_2 M^2 m_V^2\nonumber\\&& + [m_V^2 + ({\bf p} + {\bf p}')^2]({\bf p} -
       {\bf p}')^2 \}/(15 m_1^2 m_2^2 m_V^2)\nonumber\\&&+4\frac{iC_{22} g_{_{D^*D^*V}}{g'}_{_{D^*D^*V}} F(\bf
q')^2}{-(\mathbf{p} + \mathbf{p}')^2-m_{V}^2} \eta_1 \eta_2  M^2 (5 m_1^2 + 5 m_2^2 + 4 {\bf p}^2) ({\bf p}^2 + {\bf p}\cdot{\bf p}')/( 15 m_1^2 m_2^2)\nonumber\\&&+
 \frac{iC_{22}{g'}^2_{_{D^*D^*V}} F(\bf
q')^2}{-(\mathbf{p} + \mathbf{p}')^2-m_{V}^2} \{ [20 m_1^2 (2 m_2^2 + {\bf p}^2) + 4{\bf p}^2 (-2 \eta_1 \eta_2 M^2 + 5 m_2^2 + 2 {\bf p}^2)] ({\bf p} + {\bf p}')^2/(  15 m_1^2 m_2^2)\nonumber\\&& +
    2 (-6 \eta_1 \eta_2 M^2 + 5 m_1^2 + 5 m_2^2 + 2 {\bf p}^2) ({\bf p}\cdot{\bf p}'^2 -    {\bf p}^2 {\bf p}'^2)/(15 m_1^2
    m_2^2)\}
\nonumber\\
K_{23}&&=-\frac{4}{3}\frac{iC_{23} g^2_{_{D^*D^*P}} F(\bf
q)^2}{-(\mathbf{p} - \mathbf{p}')^2-m_{P}^2} \eta_1 \eta_2 M^2
({\bf p}-{\bf p}')^2
\nonumber\\
K_{24}&&=-\frac{4}{3}\frac{iC_{24} g^2_{_{D^*D^*P}} F(\bf
q')^2}{-(\mathbf{p} + \mathbf{p}')^2-m_{P}^2} \eta_1 \eta_2 M^2
({\bf p}+{\bf p}')^2
\end{eqnarray}

where  $\mathbf{q}= (\mathbf{p} - \mathbf{p}')$ and  $\mathbf{q}=
(\mathbf{p} + \mathbf{p}')$.

\end{document}